\documentstyle[preprint,aps,a4,12pt]{revtex}
\begin{document}
\draft
\hfill\vbox{\baselineskip14pt
            \hbox{\bf ETL-xx-xxx}
            \hbox{ETL Preprint 00-xxx}
            \hbox{March 2000}}
\baselineskip20pt
\vskip 0.2cm 
\begin{center}
{\Large\bf The patching of critical points using quantum group}
\end{center} 
\vskip 0.2cm 
\begin{center}
\large Sher~Alam$^{1}$,~M.~O.~Rahman$^{2}$,~M.~Ando$^{2}$,
~S.~B.~Mohamed$^{1}$ and T.~Yanagisawa$^{1}$
\end{center}
\begin{center}
$^{1}${\it Physical Science Division, ETL, Tsukuba, Ibaraki 305, Japan}\\
$^{2}${\it  GUAS \& Photon Factory, KEK, Tsukuba, Ibaraki 305, Japan}
\end{center}
\begin{center}
{\it Physical Science Division, ETL, Tsukuba, Ibaraki 305, Japan}
\end{center}
\vskip 0.2cm 
\begin{center} 
\large Abstract
\end{center}
\begin{center}
\begin{minipage}{14cm}
\baselineskip=18pt
\noindent
 Following our recent conjecture to model the phenomenona of
 antiferromagnetism and superconductivity by quantum symmetry
 groups, we discuss in the present note how to construct
 a workable scenario using this symmetry. In particular
 we propose to patch the relevant critical points. This means
 we identify fixed points, corresponding to various $k$
 or $q$ [since the two are related] and make expansion
 around these points, to control these expansion we 
 can impose gauge structure and we thus arrive at quantum
 group based gauge theory or collection of classical gauge 
 theories which represent the condensed matter system such as
 cuprates. This is different than ordinary gauge theories
 in several ways, for in ordinary field theory one has 
 well-defined critical point here the critical points are 
 not unique or simple. In short the real transition than in 
 condensed matter system is represented by collection or 
 {\it ensemble average} of the several chosen critical points 
 which come from ordinary field theory [gauge theory].
 This idea may reveal the connection between
 Hubbard model and gauge theory and string theory.
 In short it can lead 
 to the non-perturbative formualtion of Hubbard and
 other condensed matter Hamiltonians.


\end{minipage}
\end{center}
\vfill
\baselineskip=20pt
\normalsize
\newpage
\setcounter{page}{2}
	In a previous work one of us \cite{alam98} have advanced 
the conjecture that one should attempt to model the phenomena of
antiferromagnetism and superconductivity by using quantum
symmetry group. Following this conjecture to model the phenomenona of
antiferromagnetism and superconductivity by quantum symmetry
groups, three toy models were proposed \cite{alam99-1}, namely,
one based on ${\rm SO_{q}(3)}$ the other two constructed with
the ${\rm SO_{q}(4)}$ and ${\rm SO_{q}(5)}$ quantum groups. 
Possible motivations and rationale for these choices are 
were outlined. In \cite{alam99-2} a model to describe quantum 
liquids in transition from 1d to 2d dimensional crossover using 
quantum groups was outlined.    

	In this short note we turn our attention to an idea to
construct a theory based on patching critical points so as to
simulate the behavior of systems such as cuprates.
To illustrate our idea we start with an example which
has been considered by Frahm et al., \cite{fra98}. The model
deals with antiferromagnetic spin-1 chain doped with
spin-1/2 carriers. One can write the Hamiltonian as
consisting of two parts exchange and hopping
\begin{eqnarray}
H &=& \sum_{n=1}^{L} H^{exch}_{n,n+1}+H^{hopp}_{n,n+1},\nonumber\\
 H^{exch}_{i,j} &=& \frac{1}{2}(\frac{1}{S_iS_j}
{\bf S}_i \cdot {\bf S}_j-1+\delta_{_{S_iS_j,1}}
[1-({\bf S}_i\cdot{\bf S}_j)^2]),\nonumber\\
H^{hopp}_{i,j} &=& -(1-\delta_{_{S_iS_j,1}})
P_{ij}({\bf S}_i\cdot{\bf S}_j)
\label{i1}
\end{eqnarray}  
here as usual the quantity ${\bf S}_i^2=S_i(S_i+1)$ with
$S_{i}$ taking the value 1 or 1/2 and $P_{ij}$ permutes
the spins on sites $i$ and $j$.

This Hamiltonian is different from the Hamiltonian
generally thought to describe the carrier doped
Haldane system $Y_{2-x}Ca_{x}BaNiO_{5}$, namely
\begin{eqnarray}
H &=& \sum_{n=1}^{L} H^{exch}_{n,n+1}+H^{hopp}_{n,n+1},\nonumber\\
 H^{exch}_{i,j} &=& J \delta_{_{S_iS_j,1}}
{\bf S}_i\cdot{\bf S}_{j+1},\nonumber\\
H^{hopp}_{i,j} &=& -P_{ij}({\bf S}_i\cdot{\bf S}_j+1/2).
\label{i2}
\end{eqnarray} 
For no doping $x=0$ the interactions between the spin-1
$Ni^{+}$ reduce to the usual case of Heisenberg model,
however upon doping one gets a low energy doublet state
in an effective one-band Hamiltonian which can move
in the S=1 background which is caused by the mixing of
spin S=1/2 holes on the oxygen sites. The main
difference between the two Hamiltonians is the
biquadratic term. Eq.~\ref{i1} which contains
such a term gives spin-1 Takhtajan-Babujian chain for
hole doping $x=0$, thus in this undoped limit
the spectrum would be expected to be gapless, however
it is claimed in \cite{fra98} that it is possible
to reintroduce the gap in the continuum limit
where there is a field theoretical description of the model.
This theoretical description is effective
field theory, as should be noted with the following
observation which are of interest:
\begin{itemize}
\item{} It is known that in the undoped limit [$x=0$] 
one obtains the SU(2)$_{k=2}$ [which is related to
quanum group \cite{alam98,alam99-1,kak91,alam00}]
Wess-Zumino-Witten (WZW) \footnote{In literature
WZW is also called Wess-Zumino-Novikov-Witten [WZNW]
to be fair to Novikov!} with central charge
$c=3/2$. It is readily seen that this model is 
equivalent to three massless Majorana 
fermions which are a triplet under $SU(2)$.
\item{} For the other extreme case $x=1$ 
[the filled band] of the $S=1$ chain one obtains
SU(2)$_{k=1}$ in the low energy field theory
limit.
\item{} The mixed case or finite doping case
one obtains one free bosonic mode in the charge
sector, the spin sector contains a direct
sum of the $c=3/2$ and $c=1/2$ models with
different velocities! thus by doping a fourth
Majorana fermion is generated and this feature
is observed in two-channel Kondo physics 
\end{itemize} 

Keeping these points we look at the above in
the following manner:-

\begin{itemize}
\item{} The various SU(2)$_{k}$ corresponds to infrared
stable fixed points \cite{kak91,alam00}. Thus the
undoped and the fully doped case correspond to two different
fixed points. The doped case is an admixture of the two
as shown in \cite{fra98} and interpolates between
the spin S=1 and S=1/2 states. In a like manner we
propose to first label the fixed points by choosing
a symmetry such as $SO(N)_{q}$, or $SU(N)_{q}$ [ or
other Lie Groups of ones choice to describe the
condensed matter system] which correspond to 
certian $k$ level of WZW model.
\item{} In the next step since these $k$ values correspond
to particular WZW models we identify their low energy spectra.
\item{} From the low energy spectra we can go to a full
gauge theory structure.
\item{} We can next define a `partition' function of the
critical points which weighs the contributions of
the degrees of freedom which correspond to the
chosen critical points.  
\end{itemize} 

In this sense we can construct a representative
effective or smeared gauge structure for the condensed
matter system in other words a quantum group based
gauge structure. One can look at this as an expansion
about the critical points of the relevant degrees
of freedom using quantum groups.  
We note that a strong feature of quantum groups is 
that they unify classical Lie algebras and topology. 
In  general sense it is expected that quantum groups 
will lead to a deeper understanding of the concept of 
symmetry in condensed matter physics. 
 
\section*{Acknowledgments}
The Sher Alam's work is supported by the Japan Society for
for Technology [JST]. S. B. Mohamed thanks NEDO for support.

\end{document}